\documentclass[preprint, 12pt]{elsarticle}
\usepackage{graphicx}
\usepackage{amssymb}
\usepackage{amsmath}
\usepackage{url}
\usepackage[normalem]{ulem}
\usepackage{xcolor}

\newcommand{\avg}[1]{\left \langle #1  \right\rangle}

\begin{document}

\begin{frontmatter}

\title{Probabilistic model for Padel games dynamics}

\author[inst1]{Andrés Chacoma}
\author[inst1,inst2]{Orlando V. Billoni}

\affiliation[inst1]{
organization={Instituto de Física Enrique Gaviola (IFEG-CONICET)},
addressline={Ciudad Universitaria},
city={Córdoba},
postcode={5000},
country={Argentina}.
}            

\affiliation[inst2]{
organization={Facultad de Matemática, Astronomía, Física y Computación, (FAMAF-UNC)},
addressline={Ciudad Universitaria},
city={Córdoba},
postcode={5000},
country={Argentina}.
}

\begin{abstract}
This study applies complexity sciences to analyze the game of Padel. Data from 18 professional matches were collected, and the probability distributions of the total number of shots and the probability distribution of rallies' duration were analyzed. 
Based on these empirical observations and previous reports, a probabilistic model with two parameters was proposed to describe the game dynamics. One of them controls the probability of making a shot and the other probability of doing it offensively. 
The model also considers the offensive advantage of the team serving the ball. Using this model, an analytical expression for the probability distribution of the total number of shots was obtained and fit to the data. 
The results reveal that the complex dynamics of Padel can be effectively approximated as a stochastic process governed by simple probabilistic rules.
\end{abstract}


\end{frontmatter}

\section{Introduction}

In recent years, we have witnessed a significant advance in statistical analysis for sports competitions. 
With the rise of big data and the availability of advanced AI technology, researchers and analysts are now able to gather and analyze vast amounts of data from various sources to gain insights into performance, training, and strategy \cite{van2015automatically}. 
This new landscape in the area led to the development of sophisticated models and algorithms that can predict outcomes, identify key performance indicators, and provide recommendations for improvement. 
In this context, the use of statistical analysis in sports competitions has the potential to revolutionize the way teams prepare and compete, leading to better performance and ultimately, more wins.

The statistical methods mentioned above have been applied to a wide range of sports, including Football \cite{chacoma2020modeling,chacoma2021stochastic,chacoma2022complexity,loutfi2023highlighting,martin2019positional}, Volleyball \cite{chacoma2022simple,casimiro2023applying}, Basketball \cite{neiman2011reinforcement}, and Tennis \cite{shi2023effect}, among others \cite{zappala2022role,galeano2022using,buldu2021nonlinear}. 
In this paper, we aim to shed light on the game of Padel, which, despite its popularity in many countries, has received little attention in academic research \cite{
courel2019exploring, ramon2019effect, sanchez2020analysis, courel2021game}.

Padel is a racquet sport that originated in Mexico in the 1960's and has since gained popularity around the world. 
It is similar to Tennis but is played on a smaller court with walls that can be used to play the ball off of. 
The objective is to shot the ball over the net and onto the opponents' side of the court without them returning it. 
Padel can be played as singles or doubles and is often enjoyed as a social sport that can be played by people of all ages and skill levels.
Like other sports, it can be analyzed and understood from the perspective of dynamical systems. In this context, Padel presents itself as a complex system that is constantly changing, where players dynamically interact with each other and with the environment. The flow of the game, the strategies employed, individual skills, and the adaptation to changing court conditions are some of the elements that make Padel an intriguing dynamic system to study. By examining the game from this perspective, one can identify emergent patterns, causal relationships, and non-linear properties that contribute to a comprehensive understanding of the game.

The main idea of this work is to unveil underlying mechanisms of the game. 
To accomplish this, we utilized empirical data collected from elite Padel matches to investigate and characterize the dynamics of the game. Building upon this data, we developed a model that incorporates probabilistic rules to simulate the observed dynamics. By doing so, we aimed to generate a representation of the game's dynamics that aligns with the empirical evidence gathered from real-world matches.
Due to the inherent complexity of this system, we propose a simple and parsimonious model in order to intuitively understand the meaning and effect of each parameter and its relationship with the real dynamics.
In particular, with this model, we are interested in studying the probability of obtaining rallies of a given length, since a connection between this emergent and team performance has been shown in previous works \cite{courel2017game}.

\begin{table}[t!] 
\centering
\begin{tabular}{||c |c| c| c ||} 
\hline
Nº & Team 1  & Team 2 & Match \\ [1ex] 
\hline\hline
1& Spain & Argentina & WPC22, Final\\ 
2& France & Portugal & WPC22, Third Place \\ 
3& France & Spain & WPC22, Semi-finals \\ 
4& Argentina & Portugal & WPC22, Semi-finals \\ 
5& France & Belgium & WPC22, Quarter-finals\\
6& Spain & Chile & WPC22, Quarter-finals \\
7& Argentina & Paraguay & WPC22, Quarter-finals \\
8& Brazil & Portugal & WPC22, Quarter-finals  \\
9& Spain & Portugal & WPC22, Group A \\ 
10& Argentina & Spain & WPC21, Final \\ 
11& Brazil & France & WPC21, Third Place \\
12& Argentina & Brazil & WPC21, Semi-finals \\
13& Spain & France & WPC21, Semi-finals \\
14& Argentina & Paraguay & WPC21, Quarter-finals \\
15& Brazil & Chile & WPC21, Quarter-finals \\
16& Spain & Italy & WPC21, Quarter-finals \\
17& France & Uruguay & WPC21, Quarter-finals \\
18& Argentina & Uruguay & WPC21, Group B\\[1ex] 
\hline
\end{tabular}
\caption{Table showing the games visualized with the developed visualization software, including matches from the World Padel Championships held in Dubai 2022 (WPC22) and Qatar 2021 (WPC21).}
\label{ta:partidos}
\end{table}

\begin{figure}[t!]
\centering
\includegraphics[width=1.\textwidth]{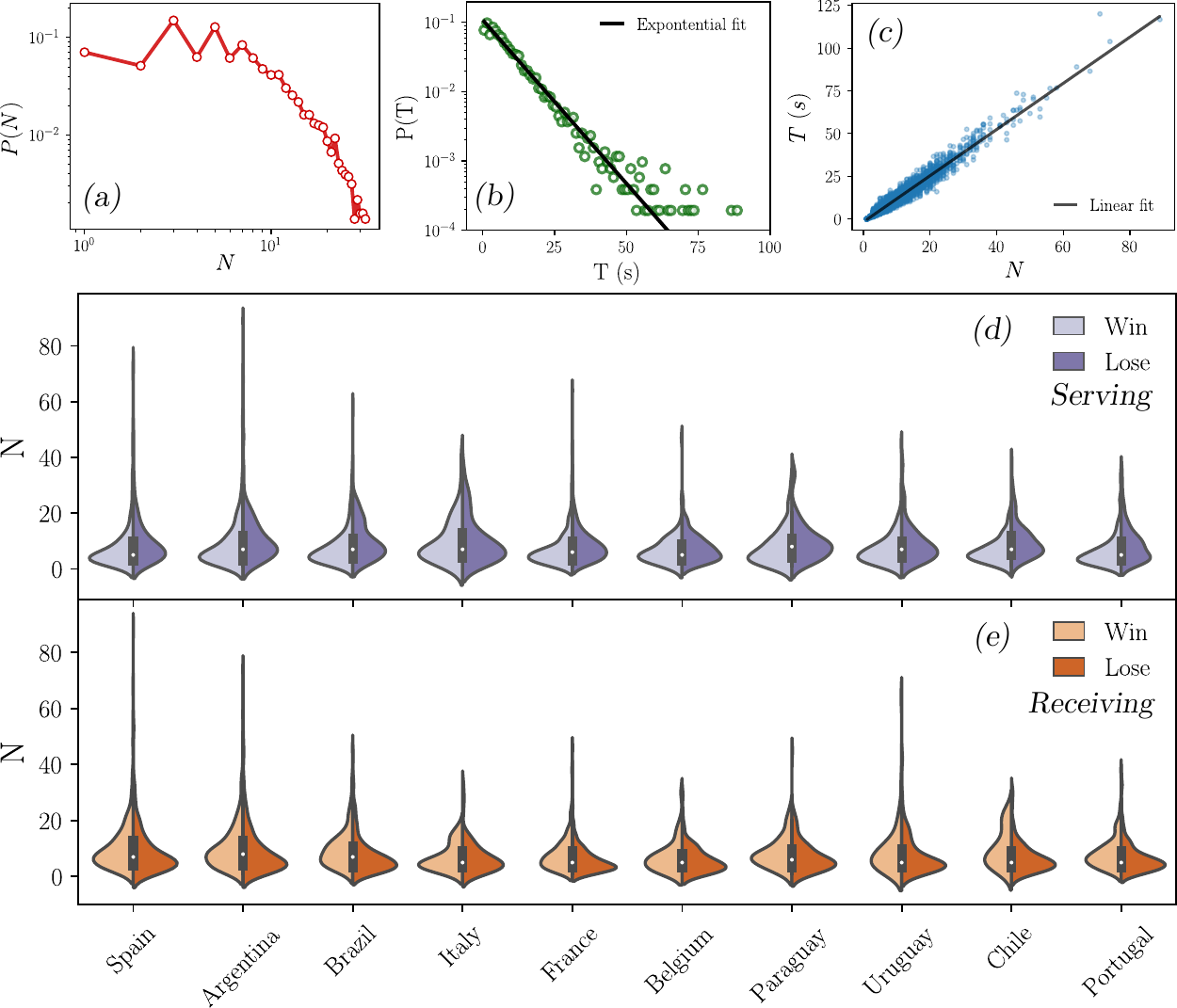}
\caption{Statistical analysis of the collected data.
(a) Probability distribution of the total number of hits in a rally.
(b) Probability distribution of the total duration of a rally.
(c) Relationship between the number of hits and the total duration of rallies.
(d) and (e) Violin plot showing the distributions of the total number of hits in the case of rallies won and lost, when the team is serving or receiving the ball, respectively.}
\label{fi:stats}
\end{figure}

\section{Data collection}

To gather the data used in this research, we developed a visualization software capable of extracting information from video recordings of Padel games. 
The software was specifically designed for a trained operator to record from each rally in the games the following,
\begin{itemize}
    \item The number of valid shots performed;
    \item The total time between the first and the last shot, i.e. the rallies' duration;
    \item The team that serves the ball;
    \item The team that wins the point.
    \end{itemize}
Table \ref{ta:partidos} provides a list of the 18 matches that were visualized. In total, we collected information from 5057 rallies. 
For transparency and reproducibility, the visualization software and an anonymized version of the collected data are available at \cite{data}.

\section{Data analysis}
\label{se:stats}

In this section, we describe some global statistical aspects observed in our data. In Fig.~\ref{fi:stats}~(a), we show the probability distribution associated with the number of shots observed in rallies, $P(N)$. We can observe a highly non-trivial behavior for low values of $N$, where the curve shows a zig-zag evolution.
This indicates the presence of singular values where the probability grows and decreases abruptly. The reader may note that the peaks are observed at odd values of $N$, and the valleys at even values. Since the team serving the ball executes the first shot, the latter tells us that serving teams have a higher probability of winning the rally in the first few shots. Likewise, the team receiving the ball has a lower probability of winning it.
In Fig.~\ref{fi:stats}~(b), we show the probability distribution of the total duration time of the rallies, $P(T)$. 
By performing a non-linear fit to the curve ($R^2\approx0.91$), we observe an exponential decay for low values of T. 
We measured the sample mean and obtained $\avg{T}=9.31$ (s).
In Fig.~\ref{fi:stats}~(c), we show the relationship between the number of shots and the total time. The observed trend was characterized with a linear fit ($R^2\approx0.98$) where we found that $T \approx \alpha N$, with $\alpha=1.35$ $(s)$. Notice, $\alpha$ indicates the mean flight time of the ball between two consecutive shots.

Finally, in the violin plots of Fig.~\ref{fi:stats}~(d) and (e), we analyzed the probability distributions of $N$ shots in a rally, for the ten analyzed teams, in the case of winning or losing the point, and with respect to whether the team is serving or receiving the ball. 
For simplicity, let us focus on describing the results for the case of Argentina.
We can see that in the case where this team is serving and wins the point, the distribution shows a marked peak for low values of $N$, while when serving and losing the point, the distribution becomes flatter and an increase of longer rallies is observed. On the contrary, in the case where the team is receiving the serve, longer rallies are observed in the cases of winning points than when losing them. From the analysis of the data associated with the rest of the teams, similar conclusions are drawn.
These observations suggest that, in general, a team's success in the rally may be closely related to its length and that the length depends, in turn, on whether the team is in possession of the serve or not.
When a team is serving, it has a high probability of winning the point if it does so quickly. 
After a certain number of shots, the advantage of having the serve decreases and the chances of winning the point become balanced for both teams.

\section{A model for Padel dynamics}

The idea behind the model is to simplify the complex dynamics of the game to a minimum. 
To achieve this, we will consider the game as a sequence of discrete events, where at each step, players must deal with a particular situation within the game.
In this frame, we can classify the universe of situations a player may encounter when executing a shot into two subsets of events: on one hand, we have shots that can be easily resolved, and on the other hand, we have shots that may pose some complications.
To streamline the discourse, henceforth we shall refer to these two subsets as the sets of defensive and offensive shots, respectively.

Within this framework, let us consider a scenario where we have two teams on the court, and at each turn, they can pass the ball to the opponent in two possible ways - through defensive shots or offensive ones.
Formally, to denote the type of shot, we use the Boolean variable S, where S=0 (S=1) indicates that a defensive (offensive) shot is performed.
We also assume that the team serving has a slight advantage over the opponent. This fact has been observed empirically in our data (see Section \ref{se:stats}) and also reported in several studies in Padel \cite{sanchez2020analysis,escudero2023influence,ramon2021comparison} and Tennis \cite{o2008importance,riddle1988probability}, and thus, we consider it essential to introduce it into the model. 
Based on this, we propose that the dynamic of the game is governed by the following rules, 

\begin{enumerate}

\item {\it Service.}
The rally starts with a team serving the ball. With probability $q$ the service is $S=1$, and with probability $(1-q)$ is $S=0$.
Considering the empirical data indicating a first serve effectiveness of approximately $90 \%$ \cite{sanchez2020analysis}, we neglect the probability of players losing their service.

\item  {\it The probability of performing a valid shot}. 
When a team is receiving a shot, we consider the following,

    \begin{enumerate}

        \item If a team is receiving an $S=0$ shot, it performs the response with probability 1.
    
        \item If a team is receiving an $S=1$ shot, it can performs the response with probability $p$. With probability $(1-p)$, the team can not produce a valid shot, missing the point.

    \end{enumerate}

\item {\it The probability of performing an offensive shot}.
When a team has to perform a shot, we consider the following,
    
    \begin{enumerate}
    
    \item If it is receiving a $S=0$ shot,  

        \begin{enumerate}
            \item If the team is receiving the service, then they can return an $S=1$ shot with probability $q$ and a $S=0$ one with probability $(1-q)$.

            \item If this shot is not from the service, then the team returns an $S=1$ shot with probability $1$.
        \end{enumerate}
    
    \item If it is receiving an $S=1$ shot, 

        \begin{enumerate}

            \item If the team is receiving the service, then they returns an $S=0$ shot with probability $1$.

            \item If this shot is not from the service, then the team can returns an $S=1$ shot with probability $q$ and a $S=0$ one with probability $(1-q)$.
            
        \end{enumerate}    
    
    \end{enumerate}

\end{enumerate}

\section{The probability of observing an N-length rally}

The probability of observing a rally with a total of $n$ shots (N-length rally), denoted by $P(N=n)$, can be calculated as follows
\begin{equation}
P(N=n) = P(S_n=1) (1-p),
\label{eq:1}
\end{equation}
where the probability that a shot at the step $n$ is offensive is denoted by~$P(S_n=1)$, and the factor $(1-p)$ indicates that the opposing team is unable to perform a shot at the step $(n+1)$, resulting in a missed point and a rally of length $n$.
As can be observed, the calculation of the probability of an offensive shot depends on the previous shot. 
The sequence of shots begins with the service, and since there is no preceding shot, we must rely on Rule 1 of the model. According to this rule, the server can perform an offensive shot with a probability of $q$, and a defensive shot with a probability of $(1-q)$. Therefore, we can calculate the probability of the first shot in a rally based on the server's chosen strategy,
\begin{equation}
\begin{split}
P(S_1=0) &= (1-q), \\
P(S_1=1) &= q.
\end{split}
\label{eq:2}
\end{equation}

For the second shot, we need to consider the following,
\begin{equation}
\begin{split}
P(S_2=0) =& P(S_2=0|S_1=0)P(S_1=0) + \\
& P(S_2=0|S_1=1)P(S_1=1), \\
P(S_2=1) =& P(S_2=1|S_1=0)P(S_1=0) + \\
& P(S_2=1|S_1=1)P(S_1=1).
\end{split}
\label{eq:3}
\end{equation}
In System~\ref{eq:3}, the first equation expresses the probability that the second shot results in $S=0$. This probability is computed by adding two terms. The first term represents the conditional probability of the second shot being $S=0$, given that the first shot was $S=0$, multiplied by the probability that the first shot was $S=0$. The second term represents the conditional probability of the second shot being $S=0$, given that the first shot was $S=1$, multiplied by the probability that the first shot was $S=1$.
Similarly, the second equation in System~\ref{eq:3} expresses the probability that the second shot results in $S=1$.

\begin{figure}[t!]
\centering
\includegraphics[width=0.9\textwidth]{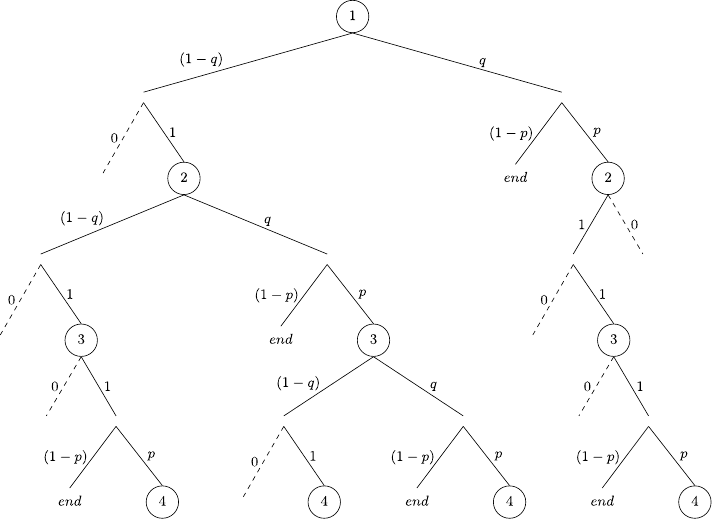}
\caption{Decision tree of the game up to shots number 4. The probability  of each shot (or {\it end}) can be evaluated multiplying the probabilities associated to the edges of the corresponding paths. Summing up the probabilities of the {\it ends} at level $n+1$ we obtain the probability of a rally of size $n$.}
\label{fi:tree}
\end{figure}

The conditional probabilities in Eq.~\ref{eq:3} can be calculated by applying the rules of the model. It is important to note that at this step, we should assume that the team is receiving the service. With this in mind, we can express the probabilities as follows,
\begin{equation}
\begin{split}
P(S_2=0|S_1=0) &= (1-q), \\
P(S_2=0|S_1=1) &= p,\\
P(S_2=1|S_1=0) &= q, \\
P(S_2=1|S_1=1) &= 0.
\end{split}
\label{eq:4}
\end{equation}
Then, by replacing Eqs.~\ref{eq:2} and \ref{eq:4} in Eq.~\ref{eq:3}, we obtain,
\begin{equation}
\begin{split}
P(S_2=0)  &= (1-q)^2 + pq , \\
P(S_2=1)  &=  q(1-q) . \\
\end{split}
\label{eq:5}
\end{equation}
For enhanced visualization of the potential paths taken by the game according to the rules of the model, Fig.~\ref{fi:tree} displays the decision tree up to shot number 4. In this schematic, nodes represent the number of executed shots, while branching links indicate whether the stroke performed was S=1 or S=0, and whether the receiving player was able to respond or not.

For the case $n>2$, the probabilities can be written as recursive sequences,
\begin{equation}
\begin{split}
P(S_{n+1}=0) =& P(S_{n+1}=0 |S_n=0 )P(S_n=0) + \\
& P(S_{n+1}=0 |S_n=1 )P(S_n=1), \\
P(S_{n+1}=1) =& P(S_{n+1}=1 |S_n=0 )P(S_n=0) + \\
& P(S_{n+1}=1 |S_n=1 )P(S_n=1).
\end{split}
\label{eq:6}
\end{equation}
Where, one more time, using the model's rules, we can calculate the conditional probabilities as follows,
\begin{equation}
\begin{split}
P(S_{n+1}=0|S_n=0) &= 0 , \\
P(S_{n+1}=0|S_n=1) &=  p(1-q),\\
P(S_{n+1}=1|S_n=0) &= 1 , \\
P(S_{n+1}=1|S_n=1) &=  pq .
\end{split}
\label{eq:7}
\end{equation}

Finally, by defining $X_n:=P(S_n=1)$ and $Y_n:=P(S_n=0)$, we can write the following system of mutually recursive linear sequences,
\begin{equation}
\left\{
\begin{aligned}
X_{n+1} &= p q X_n + Y_n,\\
Y_{n+1} &= p (1-q) X_n. 
\end{aligned}
\label{eq:8}
\right.
\end{equation}

Notice, that the roots of the characteristic polynomial related to the 2×2 matrix of system of Eqs.~\ref{eq:8},
\begin{equation}
\begin{split}
\begin{vmatrix}
p q - \lambda & 1 \\
p (1-q) & - \lambda
\end{vmatrix} = 0,\\ \\
\lambda_{1,2} = \frac{pq}{2} \pm \sqrt{ \left( \frac{pq}{2}\right) ^2 + p (1-q) }.
\end{split}
\label{eq:9}
\end{equation}
can be used to write the solution for $X_n$,
\begin{equation}
X_n = A \lambda_1^{n-1} + B \lambda_2^{n-1}.
\label{eq:10}
\end{equation}
In this frame, constants $A$ and $B$, can be calculated by using the first two results of the sequence as follows,
\begin{equation}
\begin{pmatrix}
A\\ 
B\\
\end{pmatrix} = 
\begin{pmatrix}
\lambda_1 & \lambda_2 \\ 
\lambda_1^2 & \lambda_2^2 
\end{pmatrix}^{-1}
\begin{pmatrix}
X_2 \\ 
X_3 
\end{pmatrix}.
\label{eq:11}
\end{equation}

\begin{figure}[t!]
\centering
\includegraphics[width=1.\textwidth]{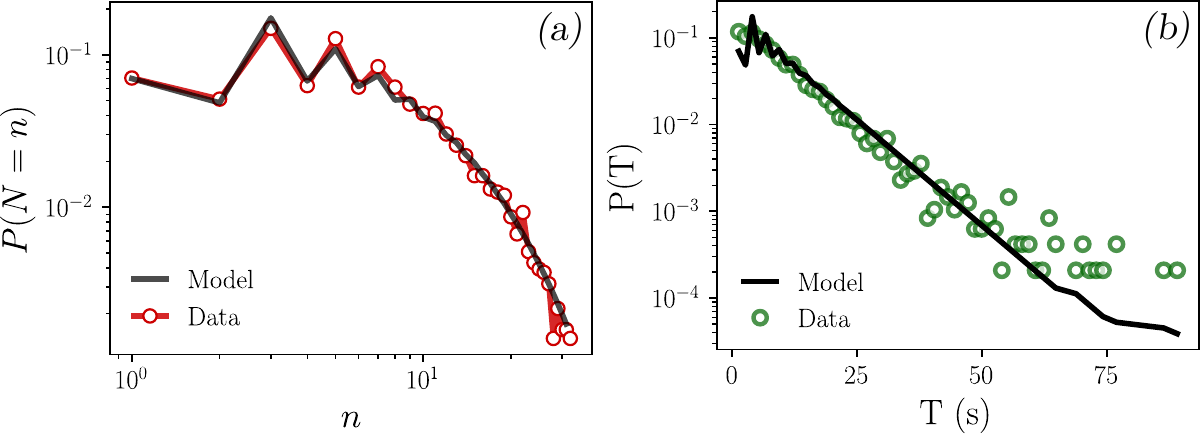}
\caption{Comparison between observed and model's probability of (a) the total number of shots in a rally, and (b) the total duration time of a rally.
The excellent agreement between the curves validates the proposed probabilistic model.}
\label{fi:Pn}
\end{figure}

In the light of the calculation described above, we have formally found the exact solution for $P(N=n)$.
In the following set of equations we summarize our result,
\begin{equation}
\left\{
\begin{aligned}
P(N=1) =& q (1-p), \\
P(N=2) =& q (1-q) (1-p), \\
P(N=n) =& X_n (1-p), \; \forall n>2.
\end{aligned}
\label{eq:12}
\right.
\end{equation}
\\
The system of Eqs.~\ref{eq:12} provides an analytical expression for the probability of observing an N-length rally. 
The next step is to fit this expression to the data and estimate the parameters $p$ and $q$ that give a good approximation of the empirical curve. 
To achieve this, we developed a minimization script that aims to minimize the Jensen-Shannon distance ($D_{JS}$) between the empirical and theoretical curves. 
The results of the minimization process are presented in Fig.~\ref{fi:Pn} (a), and we can observe an excellent agreement between the curves. 
We obtained values of $p=0.77$ and $q=0.31$, with a distance $D_{JS}\approx0.001$, indicating a small difference between curves. 
Additionally, we obtained $\avg{N}=7.65$ and $\sigma_N=5.71$ for the first and second moment in the empirical case, while for the theoretical curve, we obtained $\avg{N^{TH}}=7.57$ and $\sigma_N^{TH}=5.88$, showing a very good agreement.

In broad terms, the obtained $p$ and $q$ values indicate that (1) the probability of successfully executing a shot when receiving an offensive shot is high, and (2) the probability of executing an offensive shot is low. This aligns with the findings reported in \cite{ramon2019effect}, which compared Padel with Tennis. The study suggest that in Padel, due to the dimensions and structure of the court, the defending team has a higher likelihood of returning more shots, while the probability of hitting a winning shot is comparatively lower.
On the other hand, it is important to clarify here that, since we are integrating information from different tournaments, matches, and players, these values represent effective quantities. Note that $p$ and $q$ may vary when studying other datasets with players of different levels. In that sense, using our model to obtain the values of these probabilities in different groups could be useful for evaluating and comparing the level of the groups.

Regarding probability distribution $P(T)$, since the relationship between $N$ and $T$ can be well approximated by a linear function with a slope of $\alpha$ (see Fig.~\ref{fi:stats} panel c), the probability distribution of the rallies' total duration time can be approximated as $P(T) \approx P(\alpha N)$. 
In Fig.~\ref{fi:Pn} (b), we present a comparison between the theoretical distribution and the empirical one. Upon visual inspection, we can see a good agreement, with a measured distance between distributions of $D_{JS}\approx 0.016$.
In this case, we can calculate the theoretical mean value as $\avg{T^{TH}} = \alpha \avg{N^{TH}}= 10.22 \;(s)$.
This value differs from the sample mean by less than $1$ second  (see Section \ref{se:stats}), and it is consistent with the reported in \cite{courel2017game}.

Lastly, we can focus on describing the peaks in the distribution $P(N=n)$ for low values of $n$, which highlight the importance of having the service in the game. When a team serves, the probability of winning the point is higher at the beginning of the rally. This explains the appearance of peaks at odd values of $n$, such as $n=1,3,5$. Conversely, when a team receives the service, the chances of winning decrease, as we can see for $n=2,4,6$. Additionally, we observe that the advantage of having the service disappears after the shot $n=7$, where the peaks in the curve start to disappear.
This last observation coincides with previous studies, where it has been reported that the advantage of having the service starts to significantly decrease after the sixth and eighth shot \cite{sanchez2020analysis}.

\section{Conclusions}

In this work we show that, like many other sports, the game of Padel can be thought, modeled, and analyzed from the perspective of complexity sciences.
To study this system, first, we programmed a visualization software that allowed us to collect data from 18 elite professional matches.
With this data, we mainly focus on understanding the highly non-trivial behavior of the probability distribution of the total number of shots in the rallies, whose shape shows a zig-zag type evolution for low values of this observable.
It is interesting to mention that probability distributions of this type have been previously reported in other sports, such as in the case of volleyball \cite{chacoma2022simple}, where players have three hits to attack, and usually do it on the last one, then there is an increase in the probability of values of $n$ that are multiples of 3.
Also we observed that a team's success in a rally is related to the length of the rally where the length depends on whether the team is serving or not.
When a team is serving, it has a high probability of winning the point in a short rally. After a certain number of shots, the chances of winning the point become balanced or even worst for the serving  team.

To analyze the rally's length statistic, we have proposed a probabilistic model, where the game dynamic is governed by two parameters that control the probability of making a hit and the probability of doing it offensively. Likewise, a differentiation in the rules was proposed to contemplate the offensive advantage that the team serving the ball has.
Based on this conceptualization of the game dynamics, we were able to obtain analytically a closed mathematical expression for the probability distribution of the total number of shots in rallies. Then, we fit this expression to the data and obtained the parameters that minimize the Jensen-Shannon distance between the empirical and analytical distributions, finding an excellent agreement.

To conclude, we would like to emphasize that our proposed model does not aim to capture the entirety of the complexity inherent in the game of Padel. Rather, it serves as an initial, simple model that lays the foundation for the incorporation of concepts to explain more refined aspects of the phenomenology. 
In this regard, we consider interesting to incorporate spatiotemporal information into the model, such as the players' positions on the court at the moment of a shot or the time intervals between events. With this information, we could attempt to detect and model recurring patterns and evaluate their relationship with the game's development. Approaches with this spirit have been previously attempted with successful results in other sports \cite{galeano2022using,yamamoto2018examination}.
Additionally, it would be intriguing to integrate behavioral rules that aim to model the impact of individual decision-making. In this context, the application of game theory has demonstrated considerable efficacy in other racket sports \cite{ely2017agents,zhang2017analytical}.
Lastly, we would like to underscore that, although the game of Padel exhibits a complex dynamics, our results show that the underlying mechanisms that govern the dynamics can be mathematically represented as a stochastic process that evolves from simple probabilistic rules, which represents a significant new advance in understanding the dynamics of the game.

\section*{Acknowledgement} 

We acknowledge valuable enriching discussions with Lucía Pedraza. 
This work was partially supported by CONICET under Grant No. PIP 112 20200 101100; FonCyT under Grant Nos. PICT-2017-0973; and SeCyT-UNC (Argentina).


\end{document}